\documentstyle[epsfig,deluxetable]{mn}{}

\begin{document}
\def\mpc{h^{-1} {\rm{Mpc}}}
\def\kpc{h^{-1} {\rm{Kpc}}}
\def\up{h^{-3} {\rm{Mpc^3}}}
\def\uk{h {\rm{Mpc^{-1}}}}
\def\lsim{\mathrel{\hbox{\rlap{\hbox{\lower4pt\hbox{$\sim$}}}\hbox{$<$}}}}
\def\gsim{\mathrel{\hbox{\rlap{\hbox{\lower4pt\hbox{$\sim$}}}\hbox{$>$}}}}
\def\kms {\rm{km~s^{-1}}}
\def\masa{h^{-1}\rm{M_{\odot}}}
\def\apj {ApJ}
\def\aj {AJ}
\def\aa {A \& A}
\def\mnras {MNRAS}

\newcommand{\mincir}{\raise
-2.truept\hbox{\rlap{\hbox{$\sim$}}\raise5.truept\hbox{$<$}\ }}
\newcommand{\magcir}{\raise
-2.truept\hbox{\rlap{\hbox{$\sim$}}\raise5.truept\hbox{$>$}\ }}

\title{Richness Dependence of the Recent Evolution of Clusters of Galaxies}

\author[Plionis, Tovmassian, Andernach]
{
  \parbox[t]{\textwidth}
  {
   Manolis Plionis$^{1,2}$, Hrant M. Tovmassian$^{2}$, Heinz
   Andernach$^{3}$
   }
    \vspace*{6pt}\\ 
  \parbox[t]{15 cm}
  {
   $^1$ National Observatory of Athens, Lofos Koufou, P.Penteli 152
   36, Athens, Greece \\
   $^2$ Instituto Nacional de Astrof\'{\i}sica \'Optica y
   Electr\'onica, AP 51 y 216, 72000, Puebla, Pue, Mexico \\
   $^3$ Argelander Inst. f\"ur Astronomie, Universit\"at Bonn, D-53121
   Bonn, Germany (on leave of absence from Univ. Guanajuato, Mexico) \\
   }
}

\date{\today}

\maketitle

\begin{abstract}
We revisit the issue of the recent dynamical evolution of clusters of
galaxies using a sample of 
ACO clusters with $z<0.14$, which has been selected
such that it does not contain clusters with multiple velocity components nor strongly merging or 
interacting clusters, as revealed in X-rays.
We use as proxies of the cluster dynamical state the projected cluster
ellipticity, velocity dispersion and
X-ray luminosity. We find indications for a recent 
dynamical evolution of this cluster population, which however strongly depends on the
cluster richness. Poor clusters appear to be undergoing their primary phase of 
virialization, with their ellipticity increasing with redshift with a
rate ${\rm d}\epsilon/{\rm d} z \simeq 2.5\pm 0.4$, while the richest
clusters show an ellipticity evolution in the opposite
direction (with ${\rm d}\epsilon/{\rm d} z \simeq -1.2\pm 0.1$),
which could be due to secondary infall. When taking into account
sampling effects due to the magnitude-limited nature of the ACO
cluster catalogue we find no significant
evolution of the cluster X-ray luminosity, while the velocity
dispersion increases with decreasing redshift, independent of the
cluster richness, at a rate ${\rm d}\sigma_v/{\rm d}z \simeq -1700 \pm
400$ km s$^{-1}$.
\end{abstract}


\section{Introduction}
Structure formation in CDM models proceeds by hierarchical anisotropic accretion 
of smaller units into larger ones, along filamentary large-scale structures 
(e.g.\ Zeldovich 1970; Blumenthal et al.\ 1984; Shandarin \& Klypin 1984). 
The largest gravitationally bound, or nearly so, cosmic objects are clusters of galaxies, for which
indeed, there are indications supporting their formation by hierarchical aggregation 
of smaller systems along filaments (e.g. West, Jones, \& Forman 1995; Plionis \& Basilakos 2002). 
Since the perturbation growth rate depends on different cosmological models and the dark matter 
content of the Universe (e.g. Peebles 1980; Lahav et al.\ 1991), the present dynamical state of 
clusters of galaxies and its rate of evolution contains important cosmological information
(e.g.\ Richstone, Loeb \& Turner 1992; Evrard et al.\ 1993; Mohr et al.\ 1995; Suwa et al.\ 2003; 
Ho, Bahcall \& Bode 2006).

A variety of recent studies have attempted to characterize the morphological and dynamical 
state of groups and clusters using either optical or X-ray data (Buote \& Tsai 1995, 1996; 
Kolokotronis et al.\ 2001; Jeltema et al.\ 2005; Hashimoto et al.\ 2007, and references therein)
and thus to infer the evidence for their cosmological evolution (e.g.\ Melott, Chambers \& Miller 2001; 
Plionis 2002; Jeltema et al.\ 2005; Rahman et al.\ 2006; Hashimoto et
al.\ 2007). We can divide the various studies in those that have looked
for indications of evolution at relatively high redshifts (e.g.\
Jeltema et al.\ 2005; Hashimoto et al.\ 2007) and those that have
looked for a very recent evolution (Melott et al.\ 2001, Plionis 2002; 
Rahman et al. 2006). In both types of studies there appear
contradictory results on whether the dynamical state of clusters
evolves significantly in the distant or recent past. Melott et al.\
(2001) and Plionis (2002), using the 
projected ellipticity, $\epsilon$, as a proxy of the cluster 
dynamical state (e.g.\ Kolokotronis et al.\ 2001), found a strong
recent evolution rate with ${\rm d}\epsilon/{\rm d}z\simeq 0.7-1$ for $z\mincir 0.15$. 
This appears to be in contradiction with a similar analysis for
$z<0.31$ of Rahman et al.\ (2006) 
and with numerical N-body simulations (e.g. Floor et al.\ 2003; 2004; 
Ho et al.\ 2006) that find the recent evolution of cluster ellipticity 
to be much weaker.

Clusters of high projected ellipticity are apparently still aggregating smaller groups and 
field galaxies from their surroundings. The increase of mass concentration and 
phase-mixing during virialization will tend to sphericalize the clusters, 
increase their velocity 
dispersion, X-ray luminosity and temperature. Of course this simple picture is
highly distorted by a variety of factors, like the violent merging phase, 
strong interactions with a dense environment, cluster richness, interloper 
contamination, projection effects, etc.  
For example, the analysis of numerical simulations by Jeltema et al.\ (2008) shows 
that, using morphological criteria, less than 50\% of clusters appearing relaxed in 
projection are truly relaxed.

Therefore, in our present work we attempt to avoid systematic effects, 
as much as possible, by (a) selecting a cluster sample that is free of 
merging or strongly interacting clusters, (b) analysing the subsample
of clusters which are free of sampling effects related to the magnitude limit of the
 ACO parent cluster catalogue and (c) analysing separately clusters
of different richness. We will use as proxies of the cluster dynamical state 
its projected flatness [$f$: related to the usually used
ellipticity by $f=1/(1-\epsilon)$], X-ray temperature (k$T_x$) and luminosity
($L_x$), as well as velocity dispersion ($\sigma_v$). 
Of course, projection effects cannot easily be corrected for, a fact
which will tend to hide or reduce the amplitude of the possible correlations 
we are seeking between cluster dynamical state and redshift.

\section{Data}
For the purpose of this work we use the Abell, Corwin \& Olowin clusters
(1989, ACO in what follows) for which there are available
velocity dispersion, ellipticity and X-ray temperature or luminosity measurements.
Furthermore, we wish to concentrate mostly on the relatively slowly evolving clusters, via internal 
virialization processes, and make our analysis less prone to the complicated effects 
related to the dynamics of highly non-relaxed clusters, i.e.\ those in the state of merging,
or those with multiple components, which could be interacting strongly with their surroundings.
Note that we have chosen to use the ACO cluster catalogue
because of the extensive multiwavelength studies of the individual
ACO clusters and of the quality of the relevant data, which allows us
to identify (and exclude) merging and strongly interacting clusters
and be
confident regarding the reality or not of each of the clusters. 
This is not yet possible, at the same level, with the new
SDSS or 2dF based cluster catalogues, since individual cluster multiwavelength
studies of the these samples are not yet available (at least for the
majority of the clusters). 

To produce a ``clean'', of merging and interacting clusters, sample
we used an updated version of a compilation of cluster redshifts and
velocity dispersions (Andernach et al.\ 2005), which exploits all the available literature 
on galaxy redshifts to compile lists of galaxies in the direction of ACO clusters, 
and within a factor of four of the cluster's photometric redshift estimate. 
The 2007 version of the compilation is based on data from over 900 references and 
has $\sim$5500 cluster components for over 4000 different ACO clusters (3140 A- and 
870 S-clusters), as well as a list of $\sim$110,000 individual member redshifts in
3750 different ACO clusters.
 
Since our final sample strongly depends on the definition of a 
single component cluster in this list, 
we present some details regarding the identification of different cluster components.
The cluster velocity dispersion has been calculated by searching
initially for any relative maxima in 
the redshift distribution within the cluster area. All galaxies within $\pm 2500$ km s$^{-1}$ 
(i.e., just over three times the average $\sigma_v$ of 
$\sim 700$ km s$^{-1}$) around each relative maximum are included into a single
cluster component.
Subclumps of the same cluster which are closely located along the line of sight but
with less than 2500 km s$^{-1}$ separation in velocity, were separated into
different subclusters. Similarly, we register as different subclusters those
with a smaller velocity separation which were reported in the literature
as separated in the plane of the sky.
The velocity dispersion of the different clusters and subclusters were calculated, correcting
for measurement errors and relativistic effects, according to the prescriptions of Danese et
al.\ (1980), i.e.\ $\sigma_v=\sqrt{(\sigma_{\rm obs}^2 -\sigma_{\rm
err}^2)/(1+z)}$, where $\sigma_{\rm err}$ is the root-mean-square (r.m.s.) of the velocity errors of 
individual galaxies, or an adopted mean error if individual errors were not available. 
We consider clusters that have at least 4 measured galaxy redshifts, while cluster velocity
dispersions are considered only for those clusters that have a minimum 
of 10 measured redshifts.

Since our primary proxy for the cluster dynamical state is the cluster flatness, $f$, 
we start out from the sample of 342 ACO clusters for which Struble \& Ftaclas (1994)
compiled flatnesses from the literature. 
For details on the determination of the cluster projected shape we
point the reader to the original paper. 
Furthermore, we use a subsample of ACO clusters that excludes those showing evidences of 
strong merging or significant spatial distortions. The reason is that for such clusters 
most proxies of their dynamical state, used in our analysis (velocity dispersion, 
projected ellipticity, X-ray temperature and luminosity), are ill-defined.
To this end we identify and exclude clusters that, according to
Andernach et al.\ (2005), have multiple components in velocity space.
Furthermore and based (among others) 
on the analyses of Ledlow et al. (2003), De Filippis, Schindler \& Erben (2005), Hashimoto et al. 
(2007) and Leccardi \& Molendi (2008), we  
also exclude clusters that show multiple X-ray peaks or significantly 
distorted X-ray images, possibly implying a merging cluster (eg., A754, A1066, A1213, A1317, 
A1318, A1468, A1474, A1552, A1644, A1750, A2151, A2244, A2382, A2384, A2401, A2459, A2554, 
A3528, A3532) or for
which there is evidence for significant contamination of the X-ray measurement from 
the central AGN (eg., A2069, A2597). 
We caution the reader that our exclusion criteria may not completely clean our sample 
of significantly distorted clusters. As a test of such a residual contamination of our
sample, we repeat our analysis without excluding the previously mentioned distorted 
clusters, to find that now our results, though mostly unchanged,
become less statistically significant. 
This indicates that the possibly remaining such clusters in our sample 
would act towards reducing the significance of the intrinsic correlations.

We also imposed a minimum of 20 on the Abell galaxy count, $N_A$, which
is the number of galaxies brighter than $m_{3}+2$, taken from ACO. The
reason is that the ACO authors, different from Abell (1958), used a universal
luminosity function to correct for the background galaxies, which led to
most S-clusters having $N_A<30$, as well as some A-clusters in the overlap
zone (Table~6 of ACO).

With the above restrictions we are left with 150 clusters (including one S-cluster) 
with $z<0.14$, $N_A>20$, and with measured shape parameters. Of
these 140, 126 and 44 have velocity dispersion, X-ray luminosity and X-ray
temperature measurements, respectively.  The X-ray data have been taken
from the BAX database ({\tt webast.ast.obs-mip.fr/bax}, Sadat et al.\
2004) which offers X-ray luminosities based on H$_0$ = 50\,km\,s$^{-1}$ Mpc$^{-1}$
and $\Omega_{m}$=1.0. For three clusters the BAX redshift differed by more
than 5~percent from our (more up-to-date) redshift, so we multiplied the X-ray
luminosity in BAX with the factor $(z_{cl}/z_{BAX})^2$, where $z_{cl}$ is
the redshift from Andernach et al.\ (2005). The cluster sample used is presented in Table~1.

In the left panel of Figure~1 we present the redshift distribution of the 
cluster sample that we will analyse in this work (hashed histogram). The sample has a mean 
redshift of $\langle z \rangle \simeq 0.072$. However, since we wish to disentangle our 
analysis from effects related to the variable sampling of clusters of different richness 
at different redshifts, we divide our cluster sample into subsamples
of different richness. 

\begin{figure}
\epsfxsize=0.5\textwidth
\centerline{\epsffile{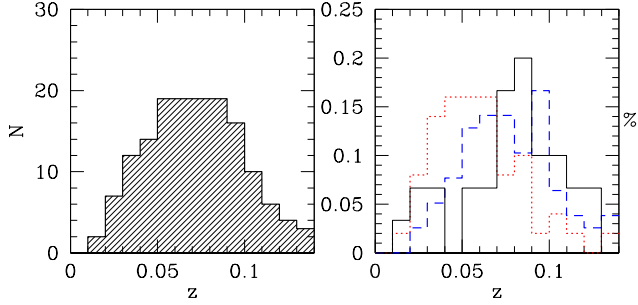}}
\caption{{\em Left Panel:} The redshift distribution of our cluster sample. 
{\em Right Panel:} The normalized redshift distribution of 3 subsamples 
based on different cluster richness. Clusters with $20<N_A<50$, $50\le N_A<80$ 
and $N_A\ge 80$ are represented by the dotted,  dashed and continuous line 
histograms, respectively.}
\end{figure}

\begin{figure}
\epsfxsize=0.5\textwidth
\centerline{\epsffile{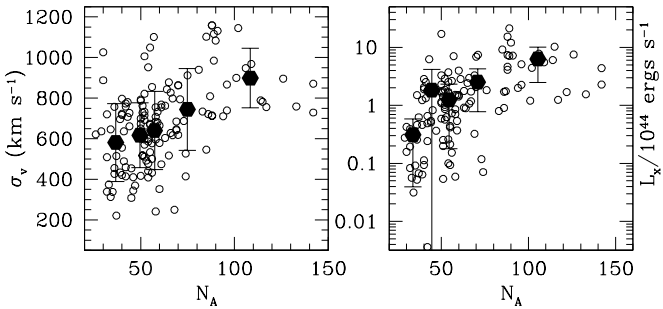}}
\caption{The correlation between Abell galaxy count, $N_A$, and the cluster 
velocity dispersion (left panel) and X-ray luminosity (right panel) for our 
sample. The filled symbols correspond to the mean values in bins of $N_A$ and 
uncertainties are $1\sigma$.}
\end{figure}

In the right panel of Figure~1 we plot the redshift distributions of clusters
in three richness classes ($20<N_A<50$, $50\le N_A<80$ and $N_A\ge 80$). 
We see that the poorer sample ($N_A<50$) has a redshift distribution significantly 
different from the richer sample ($N_A\ge80$), with mean redshifts 
of 0.059 and 0.078, respectively.

Three of the four proxies that we use for the cluster dynamical state, namely the 
cluster velocity dispersion, X-ray luminosity and temperature,  should be related 
to the total cluster mass based on the virial theorem and assuming hydrostatic equilibrium.
Indeed, we find these parameters to be strongly correlated:
the $L_x-{\rm k}T_x$, $L_x-\sigma_v$ and k$T_x-\sigma_v$ Pearson correlation coefficients 
are $R=0.75$, 0.55, 0.65, respectively, with random probabilities ${\cal P}<10^{-5}$.
Furthermore, we conjecture that cluster richness, as indicated by $N_A$, 
is proportional to the cluster total mass. We test this usual assumption by correlating
the velocity dispersion and X-ray luminosities of the clusters of our sample with $N_A$.
It is well known that the cluster X-ray luminosity is well correlated with the Abell cluster richness 
(e.g.\ Bahcall 1977; Johnson et al.\ 1983; Briel \& Henry, 1993; 
David, Forman \& Jones 1999; Ledlow et al.\ 2003), and we confirm this also for our particular 
subsample of the ACO catalogue. Correlating $N_A$ with $\sigma_v$ and $L_x$ we find the expected
strong and significant correlations, which are shown in Figure~2, with Pearson correlation coefficients 
of $R=0.46$ and 0.53, respectively, and corresponding random probabilities 
of ${\cal P}<10^{-8}$. 

\section{Results}
We revisit the issue of the morphological and dynamical evolution of clusters 
in the recent past (see Melott et al.\ 2001, Plionis 2002) using as
relevant indicators the four proxies mentioned previously. Note that
the cosmic time, within the concordance cosmological model,
corresponding to the redshift interval $0<z<0.14$ is $\sim$1.73 Gyrs, 
which is almost twice the cluster dynamical time-scale.
However,
we would like to stress that seeking indications of cluster evolution in
relatively short time-scales can be hampered by many effects among which
the intrinsic scatter of cluster shapes, the admixture of clusters of
different formation times and of 
different richness, projection effects, etc.
As one example, we would like to point out that the rate of cluster ellipticity
evolution should depend on cluster richness, since in principle massive structures 
will virialize faster than poorer ones of the same formation time. It is therefore
imperative to analyse samples of different richness
separately, and we do so further below.

As a first step, we present in Figure~3 the correlations between redshift and
the four proxies of the cluster dynamical state for the whole cluster sample. 
The continuous lines correspond to a least-squares fit to the unbinned data,
and the filled symbols correspond to the mean values in redshift bins.
We find a positive correlation, albeit quite weak, as in previous works. 
Specifically, we find Pearson correlation coefficients of $R=0.12\pm 0.02$, 0.20$\pm 0.02$ 
and 0.46$\pm 0.03$ for the $f-z$, $L_x-z$ and k$T_x-z$ correlations, respectively, with 
corresponding probabilities of being chance correlations of ${\cal P}=0.08$, 0.015 and 0.0008. 
The correlation coefficient uncertainties are estimated
by a procedure by which we exclude randomly, 100 times, 10\% of the clusters 
and re-estimate the correlation coefficient, $R$, from each reduced sample. 
\begin{figure}
\epsfxsize=0.5\textwidth
\centerline{\epsffile{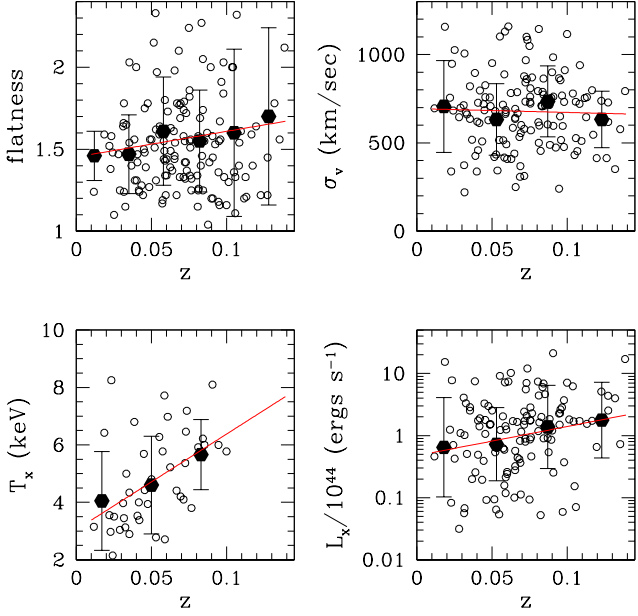}}
\caption{The apparent dependence on redshift of the cluster flatness, of the cluster velocity dispersion,
of the ICM temperature and X-ray luminosity 
for our cluster sample. The line corresponds to the best 
least-square fit to the data while the filled symbols to the mean values in redshift bins
and uncertainties are $1\sigma$. However, both the $L_x-z$ and strong k$T_x-z$ correlations are
found to be due to sample biases (see section 3.1). 
}
\end{figure}

\subsection{Accounting for systematic biases}
In general, using magnitude-limited cluster catalogues, one should 
be aware of the effects of sampling different cluster richnesses at 
different redshifts, {\em effects which could act to either weaken,
  enhance or even create apparent redshift dependent correlations}.  
Although the ACO cluster catalogue, as shown by a number of studies, 
is roughly volume-limited within $z\mincir 0.1$ (but mostly the $R\ge 1$ richness
class cluster subsample) it is essential to investigate whether
sampling biases could be disguised as ``evolutionary'' trends.
For example, as shown by analyses of cosmological simulations, richer clusters which
correspond to more massive dark matter haloes, are 
expected to be on average more elongated than poorer ones 
(e.g.\ Jing \& Suto 2002; Kasun \& Evrard 2005;
Allgood et al.\ 2006; Gottl\"ober \& Turchaninov 2006; Paz et al.\ 2006; Bett et al.\ 2007; 
Macci\'o et al.\ 2007, Ragone-Figueora \& Plionis 2007). Therefore,
the fact that at higher redshifts the sampled clusters could be typically
richer than the lower redshift counterparts (as expected in 
flux or magnitude-limited samples), implies
that we could observe an artificial correlation due to exactly the
magnitude-limited nature of the sample. Similarly, the fact that the
X-ray luminosity is correlated with the cluster richness implies 
that the average cluster $L_x$ at higher redshifts could well appear 
to be larger than the corresponding value at lower
redshifts. 

Furthermore, an
important sample incompleteness bias could also be present in the
published X-ray temperatures, since X-ray spectroscopic measurements
would be more easily available for the most X-ray brightest rather than
fainter high-$z$ clusters and therefore the apparently 
strong evolutionary trend of k$T_x$ could well be due to this
bias. Further below we test for this effect.

We now investigate the possible influence of the magnitude-limited nature
of the parent ACO cluster sample by confining our analysis within a range
of absolute magnitudes (based on the 10$^{\rm th}$ brightest cluster member) for
which there appears to be no systematic redshift-dependent sampling
effects. 
\begin{figure}
\epsfxsize=0.5\textwidth
\centerline{\epsffile{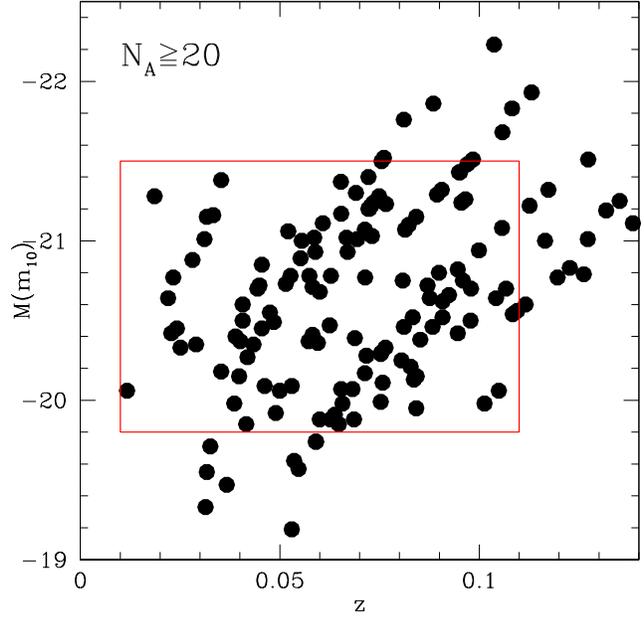}}
\caption{The dependence on redshift of the cluster absolute magnitude
  (based on the 10th brightest cluster member). In order to minimize
  sampling effects, related to the magnitude-limited nature of the ACO
  cluster sample, we investigate clusters in the reduced
  (``volume-limited'') area delineated by the continuous lines.}
\end{figure}
In Figure 4 we present the cluster $m_{10}$-based absolute magnitude as
a function of redshift for our sample. We can indeed observe the usual
redshift dependent trend which is caused by the magnitude-limited
nature of the sample.
We now use only those clusters that fall within the ``volume-limited''
area, delineated by
continuous lines ($-21.5<M<-19.8$ and $z\le 0.11$), 
for which no systematic redshift-dependent trend is
observed. We find that only the originally observed $L_x-z$ correlation
disappears, a fact that implies that this correlation is
artificial and related to the variable sampling of different cluster
richness at different redshifts. 
However, the $f-z$ and k$T_x-z$ correlations remain as significant as
for the whole sample ($R=0.16\pm 0.02$ and 0.42$\pm 0.03$, respectively, with 
corresponding probabilities of being chance correlations of ${\cal
  P}=0.05$ and 0.003), while the former correlation ($f-z$) appears to be 
even slightly stronger (although still weak in an absolute sense).

The observed $f-z$ correlation corresponds 
to a cluster ellipticity evolution rate of: 
$${\rm d}\epsilon/{\rm d} z\simeq 0.95\pm 0.18 \;,$$ 
with $\epsilon$ the projected 
ellipticity, which is in good agreement with the APM cluster results 
(${\rm d}\epsilon/{\rm d} z\simeq 0.7$) of Plionis (2002) and the study
of optical and X-ray cluster results (${\rm d}\epsilon/{\rm d} z\simeq 1$) of
Melott, Chambers \& Miller (2001). However, 
the results of Flin et al. (2004), based on Abell clusters and
analysed by Rahman et al (2006), yield a significantly lower rate of cluster 
ellipticity evolution, ${\rm d}\epsilon/{\rm d} z\simeq 0.26$.
It is interesting that N-body simulations also show a recent evolution of 
the  ellipticity of simulated clusters, but the rate of evolution is
quite low (e.g. Floor et al.\ 2004; Rahman et al.\ 2006).

We now test whether the strong k$T_x-z$ correlation could be due to
the incompleteness bias, discussed previously. To this end we compare
the $L_x-z$ correlation of those clusters that have k$T_x$ measurements
with that of the overall sample of clusters with $L_x$ data. We indeed find
that the former subsample has a strong and significant $L_x-z$
correlation ($R=0.44\pm 0.03$ with ${\cal P}=0.002$), while the parent
sample shows no significant $L_x-z$ correlation ($R=0.05$). This proves
that indeed the overall k$T_x-z$ correlation is artificial and due to
incompletness. Therefore no more reference will be given to k$T_x$
based results.

\subsection{Correlations as a function of cluster richness}
In an attempt to reconcile the different evolutionary rates of cluster
ellipticity, found in different studies, one should keep in mind the possible influence of
sampling different cluster richnesses at different redshifts (due to the magnitude-limited nature 
of the samples and of volume effects). Furthermore, 
if clusters of different richness evolve at different rates, then in comparing observations 
with simulations one should make sure to match the cluster richness (mass) distribution of the 
samples compared.
It is therefore clear that the comparison of cluster samples with a
different mix of poor and rich clusters at 
different redshifts are susceptible to interpretational error. 

We therefore analyse independently the different richness subsamples and we 
indeed find not only varying amplitudes but also opposite slopes
of these correlations.
From now on we will present results based only on the
restricted (``volume-limited'') subsample of our original cluster
sample.

In order to highlight the richness-dependent differences, we 
present below results based on the poorest and richest cluster
subsamples. For clarity we present in Fig. 5 the
selected region in the absolute magnitude - redshift plane for the
different richness subsamples.
\begin{figure}
\epsfxsize=0.5\textwidth
\centerline{\epsffile{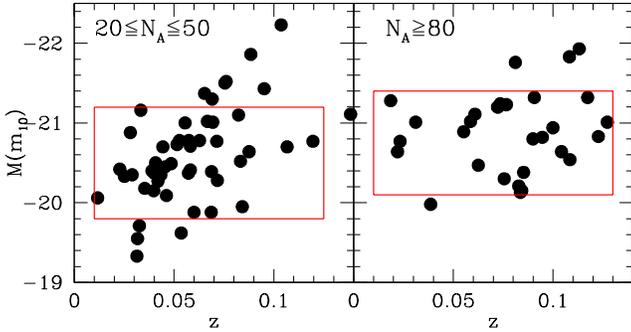}}
\caption{The dependence on redshift of the cluster absolute magnitude
  (based on the 10th brightest cluster member) for the richest and
  poorest cluster sample. The reduced
  (``volume-limited'') area used in the richness-dependent analysis is
  delineated by the continuous lines.}
\end{figure}

The Pearson correlation coefficients for the different correlations
and richness subsamples are shown in Table 2.
We find the $f-z$ correlations for the poorest and the richest cluster
subsamples to have opposite signs. They also show higher absolute amplitudes
than in the full cluster sample.
We also find a $\sigma_v-z$ correlation, in
all richness subsamples, which
is washed out in the whole cluster sample (i.e., when we do not take 
into account the different cluster richness). 
Finally, we note again that we find no significant $L_x-z$ correlation in any of
the subsamples. In Figure~6 we present only the significant correlations,
ie., the redshift dependence of the 
cluster mean ellipticity (left panel) and of the velocity dispersion (right
panel), binned in the redshift axis,
for the poorest (open points and continuous line) and the
richest (filled points and dashed line) samples, respectively.

It is important to note that the rate of ellipticity evolution for 
the poorer cluster subsample is larger than that of the whole cluster
sample, with
$${\rm d}\epsilon/{\rm d} z\simeq 2.4\pm 0.4 \;\;\;\;\;\; (20\le N_A\le 50)\;,$$
\begin{figure}
\epsfxsize=0.5\textwidth
\centerline{\epsffile{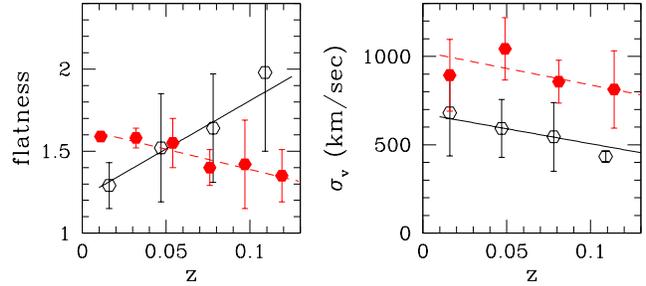}}
\caption{The redshift-flatness (left panel) and redshift-velocity
  dispersion (right panel) mean correlations for the poorest (open symbols and
  continuous lines) and the richest (solid symbols and dashed lines)
cluster subsamples.}
\end{figure}
while the corresponding rate for the richest subsample is 
$${\rm d}\epsilon/{\rm d} z\simeq -1.2\pm 0.1 \;\;\;\;
(N_A\ge 80) \;,$$
which is opposite to
the trend found for the poorest clusters, i.e., the cluster ellipticity
increases with decreasing redshift.

\begin{figure}
\epsfxsize=0.5\textwidth
\centerline{\epsffile{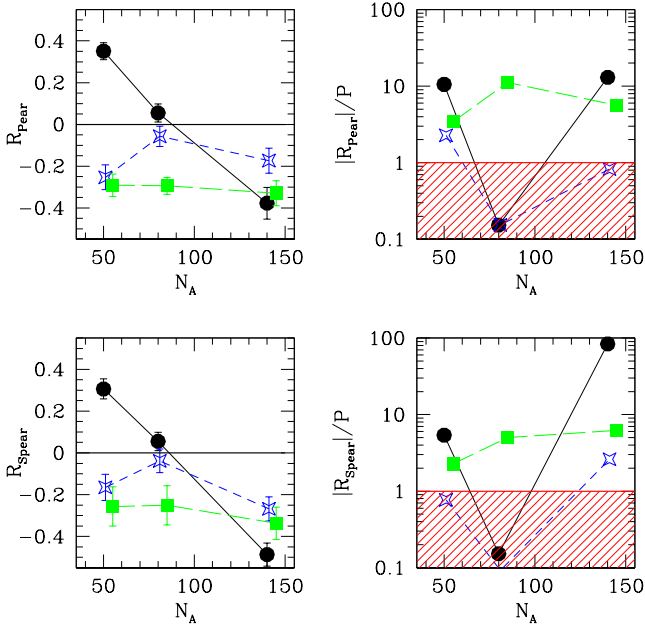}}
\caption{The correlation coefficients (left panels) and a measure of their significance (right panels) 
as a function of cluster richness (upper panels: Pearson; lower
panels: Spearman). The three (mostly) unaffected by incompletness proxies
are shown: $f-z$ (black solid line, solid circles), $L_x-z$ (blue
short-dashed line, squares) and $\sigma_{v}-z$ (green long-dashed line, solid squares),
as a function of Abell count, $N_A$. In the right panels we present an indication of significance
of the correlations with those having low values of $R/{\cal P}$ being
not significant. The shaded region corresponds to $R/{\cal P}\mincir 1$.}
\end{figure}

In order to visualize better the effect of cluster richness 
on the sign and the strength of the
correlations of the three (unbiased) proxies of the cluster dynamical-status
with redshift, we present in Fig.~7 (left panels) both the 
Pearson and Spearman correlation coefficients, evaluated in the three richness bins.
We remind the reader that positive or negative
correlation coefficients indicate that, on average, the cluster parameter decreases or 
increases towards lower redshifts, respectively.
In the right panels of Fig.~7 we present a joint indication of the
significance and strength of each 
correlation in the form of the ratio
between the correlation coefficient, $R$, and the probability, ${\cal P}$, that it is a random 
correlation. Large values of this ratio (and definitely $>1$) indicate relatively strong and
significant correlations. Different line styles and 
symbols in Fig.~7 correspond to the different cluster parameters (see figure caption). 
Correlation coefficient uncertainties are again estimated according to the procedure described 
earlier. 
The main results are:
\begin{itemize}
\item We find indications, of varying significance, for a recent
  evolution of two out of the three (unbiased) proxies of the cluster dynamical
  state (flatness and velocity dispersion).
\item There is a change of the evolutionary behavior of the cluster
 flatness as a function of richness. The correlation changes to 
anti-correlation going from poor to richer clusters. 
The intermediate richness subsample shows no $f-z$ correlation and
therefore there seems to be a smooth transition of the sign of the
$f-z$ correlation from the poorest to the richest clusters.
The rate of ellipticity evolution for 
the poorest and richest cluster subsamples are ${\rm d}\epsilon/{\rm d} 
z\simeq 2.4$ and $-1.2$, respectively.
\item The most significant evolutionary trend is that of cluster flatness with the
velocity dispersion following. The rate
of the $\sigma_v$ evolution is ${\rm d} \sigma_v/{\rm d}z\simeq
-1700\pm 400$ km s$^{-1}$, independent of the richness.
\end{itemize}

\subsection{Robustness Tests}
\subsubsection{ Does ${\rm d}{\epsilon}/{\rm d}z$ depend on limiting redshift ?}
\begin{figure}
\epsfxsize=0.5\textwidth
\centerline{\epsffile{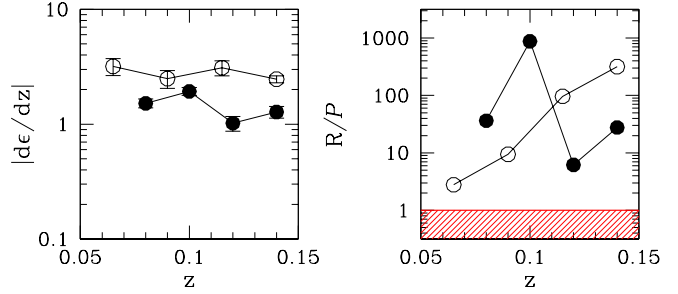}}
\caption{{\em Left Panel:} The dependence of the cluster ellipticity evolution 
rate on the limiting sample redshift. Filled and open 
points correspond to the richest and poorest samples, respectively. Note that since 
 ${\rm d}\epsilon/{\rm d}z$ is negative for the richest subsample, we plot its absolute value.
{\em Right Panel:} 
A measure of the corresponding significance of the estimated ${\rm d}\epsilon/{\rm d}z$ 
(see main text). Values corresponding to $R/{\cal P}\mincir 1$ (shaded region)
are not significant.}
\end{figure}

In order to test whether the evolution rate of cluster flatness is
sensitive to the sample limiting redshift, and thus to a few redshift
outliers, we plot in the left panel of Figure~8 
$|{\rm d}\epsilon/{\rm d}z|$ as a function of limiting sample redshift for the richest (filled 
points) and the poorest (open symbols) subsamples. The individual uncertainties are again estimated
using a procedure by which we exclude randomly, 100 times, 10\% of the clusters 
and re-estimate ${\rm d} \epsilon/{\rm d}z$ from each reduced sample. 
In the right panel of Fig.~8 we also provide the $R/{\cal P}$ indication of significance of each 
measured $|{\rm d}\epsilon/{\rm d}z|$ value.
As can be seen the amplitude of the evolutionary trend does not depend on the limiting 
redshift, while the significance of the
correlation for the poorest cluster subsample, although still
(relatively) strong, drops at lower redshifts, a fact which we 
attribute to the small number of available clusters.

\subsubsection{Are the evolutionary trends due to mass-dependent systematic effects?}
In order to demonstrate clearly that the observed evolutionary trends
are not related to any residual cluster mass-dependent systematic
effect we plot in Fig.~9 the mean flatness, $\sigma_v$ and $L_x$ for
two sets of well-separated
redshift bins and as a function of richness, ie., an analog of Fig.~6 but 
as a function of $N_A$. If there was no real evolution, one should
have expected  to see a trend of all proxies with $N_A$ (due to their
dependence on cluster mass), but
overlapping for the lower and higher-$z$ subsamples. Alternatively, if
the evolutionary trends are real we should see systematic, non-overlapping in $z$, 
offsets between the trends in different ranges of $N_A$. Indeed, as
can be seen from Fig.~9, 
the only parameter of which the richness-dependence overlaps in
redshift is $L_x$, which however we have already correctly 
identified as non-evolving with redshift.
\begin{figure}
\epsfxsize=0.5\textwidth
\centerline{\epsffile{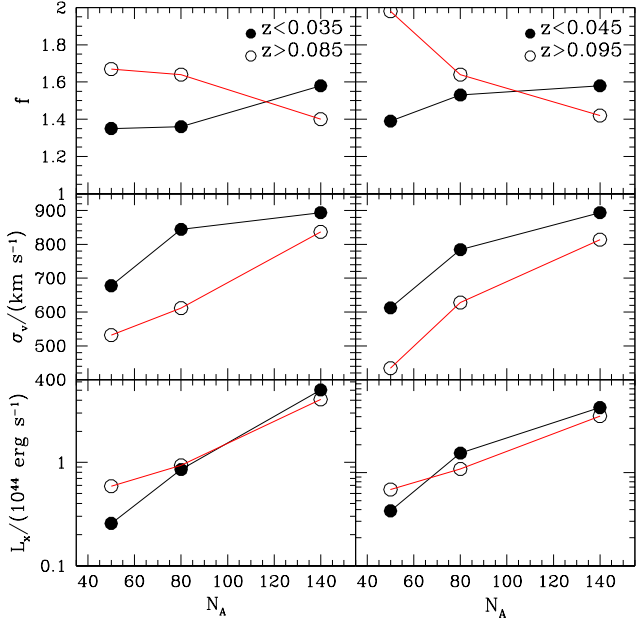}}
\caption{The variation with richness of the mean cluster dynamics
  proxies for two sets of well separated redshift-bins
  (indicated in the plot).}
\end{figure}

\subsection{Possible Interpretation}
These results can be interpreted if the population of poorer clusters is dynamically younger 
than that of the richer ones, and that they are now going through their primary virialization 
process, which tends to sphericalize their original anisotropic morphologies.

Regarding the rich clusters, one could have interpreted the fact that their 
velocity dispersion increases at lower redshifts again as an indication 
of them becoming more virialized, since once the cluster potential 
has accumulated the bulk of the mass, via infall and merging,
then the virialization processes will tend to increase the velocity
dispersion.
However, if this were the case then there should have been also signs
of the clusters becoming more spherical at lower redshifts, which is exactly the opposite than what
is observed.
Therefore, the previous interpretation does not seem plausible. Rather it appears that the
rich clusters of our sample have already reached a virialized state, while the 
redshift dependent changes in their dynamical state (evidenced by the
increase of their flatness and velocity dispersion) are probably caused by secondary infall 
(Gunn  1977; see also Ascasibar, Hoffman \& Gottl\"ober 2007 and references therein; 
Diemand \& Kuhlen 2008).

If on the other hand the poor cluster population is currently going
through the primary virialization process, 
there should be a clear correlation between cluster flatness and velocity dispersion, as well as 
with ICM X-ray luminosity.
Since we have taken good care to exclude multiple component and merging clusters, 
we believe that the velocity dispersion measurement 
is not significantly contaminated by the infall component of the merging process 
or by strong tidal effects
and thus it should indeed reflect the cluster DM gravitational potential.
Similarly, the ICM (traced by the X-ray luminosity) should not be significantly 
contaminated by effects related to shocks induced during the merging 
process and thus it should also reflect the dynamical state of our clusters.

\begin{figure}
\epsfxsize=0.5\textwidth
\centerline{\epsffile{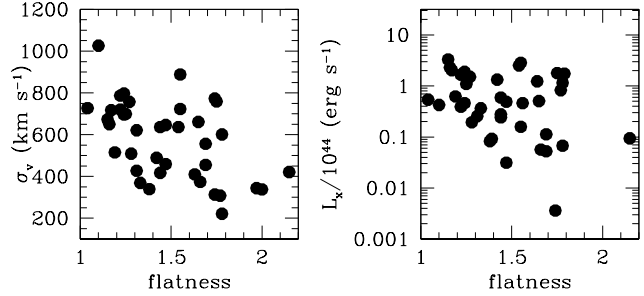}}
\caption{Correlation of two proxies for cluster dynamical state for the poorest subsample
($20<N_A<50$, $z<0.1$).
{\em Left:} The dependence of cluster velocity dispersion on cluster flatness. 
{\em Right:}  The dependence of cluster X-ray luminosity on cluster flatness.}
\end{figure}

We therefore correlate, for our poorest cluster subsample, cluster flatness with 
velocity dispersion and X-ray luminosity. Since we are
not seeking evolutionary trends we do not impose limits in absolute
magnitude. We find a strong and significant 
anti-correlation in the first two cases (Figure~10). The Pearson correlation coefficients are 
$R=-0.43\pm 0.03$ and $-0.39\pm 0.03$, for the $f-\sigma_v$ and $f-L_x$ 
correlations respectively, with corresponding random probabilities of 
${\cal P}=3\times 10^{-3}$ and 0.007. 
These results indeed show the expected behavior for a cluster population at different stages 
of virialization.
It is important to note that similar correlations are not found in the richer subsamples, 
as expected if these clusters are already virialized. 

We conclude that  poor and relatively nearby clusters are currently evolving dynamically 
and they appear to be at various stages of virialization. 
Richer clusters (at the redshift range probed) are probably already virialized, but
show indications of being affected by secondary infall.

\section{Discussion \& Conclusions}
Our current analysis supports previous results regarding the recent ($z\mincir 0.14$) evolution of 
the ellipticity and dynamics of clusters of galaxies. We have found however that the direction 
of evolution is different for clusters of different richness. Regarding the rate of ellipticity
evolution we find ${\rm d}\epsilon/{\rm d} z\simeq 0.95$ for our full cluster 
sample, which is in good agreement 
with Melott et al. (2001) and Plionis (2002), but in disagreement with Rahman 
et al.\ (2006) who quote a value ${\rm d}\epsilon/{\rm d} z\simeq 0.2$. 
It is important to note that the evolution rates for the poorest and richest of our clusters have 
opposite signs: 
${\rm d}\epsilon/{\rm d} z\simeq 2.4\pm 0.4$ 
and ${\rm d}\epsilon/{\rm d} z\simeq -1.2\pm 0.1$, respectively.
It is clear that the overall evolution rate of a sample of clusters depends on the
richness mix, and this could well be the reason why different studies find different values of 
${\rm d}\epsilon/{\rm d}z$.

Summarizing, we would like to point out that:

1. From the observational point of view, the relatively strong recent evolution of cluster
ellipticity and dynamical state applies mostly to poor
clusters, for which the rate of evolution (${\rm d}\epsilon/{\rm d} z\simeq 2.4$)
is significantly larger than that of the whole sample put together 
(${\rm d}\epsilon/{\rm d} z\simeq 0.95$). Rich clusters appear to have 
reached an equilibrium state earlier 
and thus they do not show signs of positive evolution in the recent 
past, but rather of a negative evolution (${\rm d}\epsilon/{\rm d} z\simeq -1.2$), 
possibly due to secondary infall (eg., Gunn  1977; Ascasibar, Hoffman \& Gottl\"ober 2007, 
Diemand \& Kuhlen 2008).
There are also indications for a recent evolution of the cluster
velocity dispersion, increasing with decreasing redshift but apparently
independent of the cluster richness, with a rate ${\rm d}\sigma_v/{\rm d}z \simeq -1700 \pm
400$ km s$^{-1}$. No evolution is observed of the ICM X-ray luminosity.

2. The discrepancy with the ellipticity evolution results of Flin et al. (2004),
analysed in Rahman et al. (2006),
could well be due to the latter study not taking into account the cluster richness 
dependence of the effect, or due to not excluding strongly interacting
and merging clusters
and possibly also to the sample's larger limiting redshift ($z\mincir 0.31$).

3. The discrepancy with N-body results could be due to a number of reasons. 
A quite probable reason is related to the fact that
the simulated clusters are predominantly rich (Floor et al. 2003; 2004) for which, 
as we have shown, there is no observational evidence for a recent 
positive evolution, but rather there are indications for a mild negative evolution.
Rahman et al. (2006) simulated also poorer clusters,
but the total number of analysed clusters is quite small ($N=41$). Since the 
intrinsic scatter of cluster 
(and halo) shapes is large and the observed effect appears to be inherently weak, 
a large cluster sample is probably necessary in order to clearly establish the evolutionary effect.
Furthermore, if the richness mix of the simulated clusters is significantly different from that 
of the observed clusters (or between different observational cluster samples), then due to the
richness dependence of the effect, one could derive different rates of evolution from different 
richness mixes.

\section*{Acknowledgments}
This research has made use of the X-Rays Clusters Database (BAX)
which is operated by the Laboratoire d'Astrophysique de Tarbes-Toulouse (LATT),
under contract with the Centre National d'Etudes Spatiales (CNES).
M.P.\ acknowledges financial support under CONACyT grant 2005-49878.
H.A.\ has benefitted from financial support under CONACyT grant
50921-F. We also thank the anonymous referee for valuable suggestions.

\newpage

\begin{table}

\caption{The Abell cluster sample used.}
\tabcolsep 4pt
\begin{tabular}{cccccccc} \\ \hline
ACO   & $N_A$   & $z_{LG}$  & $N_z$ & $\sigma_v$ & $f$  & $L_x$ $^a$  & k$T_x$   \\ 
\multicolumn{4}{c}{} & (km s$^{-1}$)   &   & $10^{44}$\,erg/s  & (keV) \\ \hline

A~~13 &  96 & 0.0946 &  39 &  867 &   1.18 &   2.26    &  6.0    \\
A~~14 &  29 & 0.0653 &  46 &  636 &   1.44 &   0.277   &         \\
A~~16 &  86 & 0.0843 &   7 &      &   1.25 &   0.900   &         \\
A~~21 &  56 & 0.0955 &  15 &  855 &   1.67 &   2.64    &         \\
A~~23 &  45 & 0.1067 &  29 &  454 &   2.32 &           &         \\
A~~27 &  46 & 0.0536 &  14 &  344 &   1.97 &           &         \\
A~~76 &  42 & 0.0407 &  13 &  459 &   1.47 &   0.490   &  1.50   \\
A~~77 &  50 & 0.0717 &   4 &      &   1.79 &   1.74    &         \\
A~~84 &  76 & 0.1013 &   9 &      &   1.30 &   1.83    &         \\
A~~95 &  52 & 0.1095 &  23 &  511 &   1.61 &           &         \\
A~112 &  50 & 0.1385 &  28 &  793 &   2.12 &  17.1$^b$ &         \\
A~114 &  30 & 0.0582 &  43 &  888 &   1.55 &   0.158   &         \\
A~119 &  69 & 0.0449 & 239 &  685 &   1.50 &   3.30    &  5.69   \\
A~126 &  51 & 0.0547 &  14 &  516 &   2.81 &   0.054   &         \\
A~147 &  32 & 0.0444 &  31 &  720 &   1.22 &   0.391   &         \\
A~150 &  55 & 0.0591 &  17 &  674 &   1.34 &   0.213   &         \\
A~193 &  58 & 0.0490 &  75 &  708 &   1.56 &   1.53    &         \\
A~195 &  32 & 0.0434 &  13 &  509 &   1.28 &   0.194   &         \\
A~260 &  51 & 0.0367 &  51 &  518 &   1.40 &   0.199   &         \\
A~272 &  52 & 0.0883 &  17 &  715 &   2.40 &   1.85    &         \\
A~367 & 101 & 0.0899 &  33 &  900 &   1.20 &   1.20    &         \\
A~376 &  36 & 0.0485 &  79 &  757 &   1.27 &   1.50    &         \\
A~389 & 133 & 0.1131 &  55 &  759 &   1.21 &   1.55    &         \\
A~399 &  57 & 0.0729 & 170 & 1101 &   1.42 &   7.06    &  6.46   \\
A~400 &  58 & 0.0242 & 125 &  683 &   1.48 &   0.706   &  2.15   \\
A~401 &  90 & 0.0735 & 170 & 1083 &   1.37 &  12.1     &  7.19   \\
A~415 &  67 & 0.0808 &  12 &  617 &   1.15 &   1.18    &         \\
A~426 &  88 & 0.0186 & 190 & 1158 &   1.59 &  15.3     &  6.42   \\
A~496 &  50 & 0.0326 & 358 &  673 &   1.15 &   3.31    &  3.13   \\
A~505 &  39 & 0.0555 &   4 &      &   1.64 &   1.23    &         \\
A~539 &  50 & 0.0290 & 159 &  699 &   1.25 &   1.10    &  3.04   \\
A~568 &  36 & 0.0761 &   5 &      &   1.56 &   0.460   &         \\
A~655 & 142 & 0.1272 &  61 &  729 &   1.22 &   4.39    &         \\
A~724 &  61 & 0.0924 &  72 &  474 &   1.53 &           &         \\
A~727 &  65 & 0.0959 &  63 &  517 &   1.56 &           &         \\
A~779 &  32 & 0.0228 &  81 &  339 &   1.38 &   0.083   &  2.97   \\
A~838 &  40 & 0.0515 &  11 &  421 &   2.15 &   0.091   &         \\
A~858 &  44 & 0.0876 &  40 &  727 &   1.04 &   0.539   &         \\
A~879 &  61 & 0.1116 &  35 &  754 &   1.20 &           &         \\
A~979 &  39 & 0.0527 &  18 &  434 &   2.33 &   0.064   &         \\
A~999 &  33 & 0.0314 &  51 &  374 &   1.66 &   0.056   &         \\
A1016 &  37 & 0.0317 &  46 &  221 &   1.78 &   0.067   &         \\
A1033 &  96 & 0.1227 &  38 &  739 &   1.22 &   5.12    &         \\
A1060 &  39 & 0.0117 & 330 &  696 &   1.24 &   0.461$^b$ &  3.15   \\
A1100 &  35 & 0.0462 &   4 &      &   1.39 &   0.092   &         \\
A1139 &  36 & 0.0389 & 152 &  427 &   1.31 &   0.256   &         \\
A1149 &  34 & 0.0714 &  49 &  313 &   1.74 &           &         \\
A1168 &  52 & 0.0908 &  46 &  597 &   1.45 &           &         \\
A1169 &  73 & 0.0590 & 106 &  687 &   1.43 &   0.119   &         \\
A1173 &  52 & 0.0758 &  56 &  571 &   1.51 &   0.937   &         \\
A1187 &  55 & 0.0749 &  16 & 1049 &   2.24 &   0.093   &         \\
A1190 &  87 & 0.0755 &  23 &  809 &   1.34 &   1.75    &         \\
A1205 &  63 & 0.0753 &  76 &  762 &   1.86 &   1.77    &         \\
A1225 &  43 & 0.1037 &  56 &  780 &   2.00 &   6.23    &         \\
A1235 & 122 & 0.1042 &   4 &      &   2.00 &   1.70    &         \\
A1270 &  40 & 0.0691 &  57 &  556 &   1.69 &   0.113   &         \\
\end{tabular}
\end{table}

\begin{table}
\tabletypesize{\scriptsize} 
 \contcaption{}
\tabcolsep 4pt
\begin{tabular}{cccccccc} \\ \hline
Cluster    & $N_A$   & $z_{LG}$  & $N_z$ & $\sigma_v$ & $f$  & $L_x$ $^a$  & k$T_x$   \\ 
\multicolumn{4}{c}{} & (km s$^{-1}$)   &   & $10^{44}$\,erg/s  & (keV) \\ \hline
A1307 &  71 & 0.0805 &  97 &  794 &   1.34 &   7.43    &         \\
A1314 &  44 & 0.0333 & 107 &  661 &   1.65 &   0.506   &  5.00   \\
A1324 &  58 & 0.0946 &  12 &  241 &   1.36 &           &         \\
A1341 &  56 & 0.1049 &  28 &  432 &   1.36 &   0.153   &         \\
A1344 &  51 & 0.0765 &   7 &      &   1.25 &           &         \\
A1346 &  59 & 0.0979 &  97 &  732 &   2.03 &   0.371   &         \\
A1364 &  74 & 0.1058 &  64 &  527 &   1.33 &   0.071   &         \\
A1367 & 117 & 0.0220 & 283 &  756 &   1.52 &   1.25    &  3.55   \\
A1371 &  55 & 0.0682 &  73 &  534 &   2.10 &   0.416   &         \\
A1377 &  59 & 0.0521 &  78 &  680 &   1.51 &   0.540   &         \\
A1380 &  76 & 0.1057 &  38 &  733 &   1.11 &           &         \\
A1383 &  54 & 0.0600 &  78 &  409 &   1.34 &   0.260   &         \\
A1407 &  56 & 0.1352 &  27 &  578 &   1.56 &   0.730   &         \\
A1412 &  86 & 0.1082 &  11 &  721 &   1.40 &   1.72$^b$ &         \\
A1424 &  52 & 0.0753 &  98 &  732 &   1.82 &   0.866   &         \\
A1436 &  69 & 0.0654 &  79 &  642 &   1.98 &   0.322   &         \\
A1448 &  70 & 0.1273 &  15 &  619 &   1.45 &   2.54    &         \\
A1452 &  46 & 0.0628 &  26 &  410 &   1.63 &           &         \\
A1496 &  58 & 0.0970 &  75 &  596 &   3.55 &   0.059   &         \\
A1520 &  45 & 0.0686 &  11 &  308 &   1.77 &   0.825   &         \\
A1541 &  58 & 0.0894 &  77 &  755 &   2.00 &   0.854   &         \\
A1620 &  42 & 0.0842 & 107 &  773 &   1.74 &   0.0036  &         \\
A1630 &  54 & 0.0648 &  34 &  440 &   1.61 &   0.100   &         \\
A1650 & 114 & 0.0836 & 220 &  789 &   1.46 &   6.05    &  5.68   \\
A1651 &  70 & 0.0842 & 228 &  864 &   2.05 &   6.92    &  6.22   \\
A1656 & 106 & 0.0233 & 794 &  948 &   1.58 &   7.77    &  8.25   \\
A1668 &  54 & 0.0638 &  48 &  586 &   1.37 &   1.71    &         \\
A1691 &  64 & 0.0722 & 111 &  843 &   1.42 &   0.889   &         \\
A1738 &  85 & 0.1173 &  59 &  546 &   1.40 &           &         \\
A1764 &  42 & 0.1196 &  12 &  414 &   1.64 &   0.656   &         \\
A1767 &  65 & 0.0713 & 159 &  878 &   1.58 &   2.43    &  4.10   \\
A1783 &  47 & 0.0688 &  57 &  369 &   1.33 &   0.364   &         \\
A1784 &  74 & 0.1262 &  12 &  414 &   2.93 &           &         \\
A1795 & 115 & 0.0625 & 127 &  782 &   1.39 &  10.3     &  5.22   \\
A1800 &  40 & 0.0755 &  91 &  723 &   1.55 &   2.85    &         \\
A1827 &  68 & 0.0657 &  10 &  250 &   1.54 &           &         \\
A1828 &  59 & 0.0627 &   6 &      &   1.49 &           &         \\
A1837 &  50 & 0.0694 &  50 &  601 &   1.78 &   1.15    &  4.20   \\
A1890 &  37 & 0.0574 &  93 &  515 &   1.19 &   0.623   &  5.77   \\
A1904 &  83 & 0.0722 & 137 &  734 &   1.55 &   0.798   &         \\
A1913 &  53 & 0.0530 &  17 &  631 &   1.62 &   0.628   &  2.78   \\
A1927 &  50 & 0.0952 &  50 &  650 &   1.16 &   2.30    &         \\
A1930 &  60 & 0.1318 &  16 &  352 &   1.78 &   3.99    &         \\
A1986 &  67 & 0.1165 &  12 &  798 &   1.59 &   1.73    &         \\
A1991 &  60 & 0.0589 &  65 &  665 &   1.80 &   1.42    &  2.71   \\
A2022 &  50 & 0.0582 &  26 &  417 &   1.44 &   0.591   &         \\
A2026 &  51 & 0.0908 &  60 &  762 &   1.20 &   0.253   &         \\
A2048 &  75 & 0.0984 &  74 &  912 &   1.33 &           &         \\
A2052 &  41 & 0.0353 &  92 &  636 &   1.54 &   2.52    &  2.89   \\
A2062 &  69 & 0.1126 &  57 &  646 &   1.33 &           &         \\
A2065 & 109 & 0.0724 &  42 &  968 &   1.32 &   5.55    &  5.37   \\
A2089 &  70 & 0.0731 &  78 &  862 &   1.72 &   2.07    &         \\
A2092 &  55 & 0.0670 &  44 &  668 &   1.78 &   0.440   &         \\
A2107 &  51 & 0.0416 &  90 &  613 &   1.25 &   1.41    &  4.00   \\
A2110 &  54 & 0.0978 &  46 &  477 &   1.84 &   3.70    &         \\
A2124 &  50 & 0.0667 & 118 &  787 &   1.22 &   1.66    &  4.41   \\
A2142 &  89 & 0.0906 & 240 &  985 &   1.58 &  21.2     &  8.10   \\
A2147 &  52 & 0.0353 &  93 &  821 &   2.03 &   2.87    &  4.34   \\
A2148 &  41 & 0.0885 &  47 &  489 &   1.42 &   1.33    &         \\
\end{tabular}
\end{table}

\begin{table}
\tabletypesize{\scriptsize} 
 \contcaption{}
\tabcolsep 4pt
\begin{tabular}{cccccccc} \\ \hline
Cluster    & $N_A$   & $z_{LG}$  & $N_z$ & $\sigma_v$ & $f$  & $L_x$ $^a$  & k$T_x$   \\ 
\multicolumn{4}{c}{} & (km s$^{-1}$)   &   & $10^{44}$\,erg/s  & (keV) \\ \hline
A2175 &  61 & 0.0965 &  84 &  768 &   1.91 &   2.84    &         \\
A2199 &  88 & 0.0311 & 471 &  714 &   1.64 &   4.09    &  3.97   \\
A2244 &  89 & 0.0999 & 116 & 1116 &   1.16 &   7.13    &  5.77   \\
A2245 &  63 & 0.0870 &  73 &  604 &   1.60 &   0.915   &         \\
A2247 &  35 & 0.0398 &  22 &  338 &   2.00 &           &         \\
A2250 &  52 & 0.0654 &  18 &  693 &   1.46 &   0.569   &         \\
A2255 & 102 & 0.0811 & 213 & 1145 &   1.25 &   4.43    &  5.92   \\
A2256 &  88 & 0.0608 & 329 & 1159 &   1.73 &   7.40    &  6.98   \\
A2372 &  42 & 0.0600 &   7 &      &   1.44 &   0.242   &         \\
A2377 &  94 & 0.0828 &  20 &  711 &   1.40 &   1.99    &         \\
A2410 &  54 & 0.0814 &  15 &  599 &   1.26 &   1.63    &         \\
A2415 &  40 & 0.0572 &  12 &  717 &   1.17 &   2.04    &         \\
A2420 &  88 & 0.0852 &  11 &  712 &   1.55 &   4.64    & 6.0     \\
A2448 &  36 & 0.0823 &  43 &  455 &   1.69 &   0.052   &         \\
A2457 &  53 & 0.0595 &  34 &  492 &   1.22 &   1.25    &         \\
A2529 &  81 & 0.1084 &  25 &  940 &   1.54 &           &         \\
A2569 &  56 & 0.0811 &  42 &  501 &   1.58 &           &         \\
A2589 &  40 & 0.0419 &  70 &  797 &   1.24 &   1.90    &  3.38   \\
A2597 &  43 & 0.0833 &  45 &  707 &   2.27 &   6.62$^c$&  3.67$^c$ \\
A2634 &  52 & 0.0317 & 254 & 1006 &   1.24 &   1.02    &  3.45   \\
A2637 &  60 & 0.0713 &  11 &  579 &   1.33 &   1.50    &         \\
A2657 &  51 & 0.0407 &  76 &  728 &   1.42 &   1.75    &  3.53   \\
A2666 &  34 & 0.0281 &  79 &  646 &   1.47 &   0.031   &         \\
A2670 & 142 & 0.0766 & 265 &  871 &   1.32 &   2.28    &  3.80   \\
A2686 &  61 & 0.0530 &   4 &      &   1.31 &           &         \\
A2700 &  59 & 0.0949 &   9 &      &   1.55 &   1.50    &         \\
A2877 &  30 & 0.0251 & 170 & 1026 &   1.10 &   0.42    &  3.50   \\
A3266 &  91 & 0.0586 & 317 & 1131 &   1.49 &   7.22    &  7.72   \\
A3376 &  42 & 0.0455 & 113 &  759 &   1.75 &   1.78    &  4.43   \\
A3395 &  54 & 0.0500 & 185 &  952 &   1.56 &   2.54    &  4.80   \\
A3571 & 126 & 0.0386 & 171 &  896 &   1.56 &   7.51    &  6.80   \\
A3667 &  85 & 0.0552 & 231 & 1102 &   1.59 &   9.16    &  6.28   \\
A3716 &  66 & 0.0455 & 216 &  827 &   1.48 &   1.06    &         \\
A4059 &  66 & 0.0475 &  45 &  628 &   1.78 &   2.98    &  3.94   \\
S~463 &  26 & 0.0401 &  99 &  621 &   1.31 &           &         \\


\tablenotetext{a}{$H_{0}=50$ km sec$^{-1}$ Mpc$^{-1}$ is used}
\tablenotetext{b}{$L_x$ corrected by a factor $(z_{cl}/z_{BAX})^2$}
\tablenotetext{c}{$L_x$ and k$T_x$ not used due to a possible 
significant central AGN contamination; McNamara, et al. (2001); Morris \& Fabian (2005).}

  \end{tabular}
\end{table}

\begin{table}
\caption[]{Pearson correlation coefficients and their significance
for the correlations with redshift of the 
three (unbiased) proxies of the cluster dynamical state. Results are
based on the ``volume-limited'' subsamples. We indicate the most
significant correlations in bold font.}
\tabcolsep 4pt
\begin{tabular}{ccccccccccccc} \\ \hline
 $N_A$ & \multicolumn{3}{c}{$f-z$} &\multicolumn{3}{c}{$\sigma_v-z$} & \multicolumn{3}{c}{$L_x-z$} \\ \hline
 
          & $\#$ &R & ${\cal P}$ &  $\#$ &R & ${\cal P}$ &  $\#$ &R & ${\cal P}$ \\
 $>20$    & 120& {\bf 0.17} & {\bf 0.03} & 110 &  -0.05 & 0.30 & 103 & 0.06 & 0.3 \\ \\
 20-50    & 38 & {\bf 0.40} & {\bf 0.01} & 34 & {\bf -0.22} & {\bf 0.10} & 32 & -0.05 & 0.4 \\ 
 51-79    & 45 &       0.06  &       0.40  & 44 & {\bf -0.29} & {\bf 0.03} & 37 &-0.06 & 0.4 \\
 $\ge 80$ & 26 & {\bf -0.38}& {\bf 0.03} & 25 & {\bf -0.30} & {\bf 0.07} & 24 & -0.17 & 0.2 \\ \hline
\end{tabular}

%
\end{table}

\end{document}